\documentclass{icrc29}
\usepackage{graphicx,amssymb,amsmath,times}
\begin{document}
\title[TeV $\gamma$-Ray Observations of the Perseus Galaxy
Cluster]{TeV $\gamma$-Ray Observations of the Perseus Galaxy Cluster}
\author[J.S. Perkins et al.] {J.S. Perkins$^a$ and the VERITAS
  Collaboration$^b$ \\\textsf{}
  (a) Washington University in St. Louis, St. Louis, MO 63130 USA ~ \\
  (b) For full author list, see J. Holder's paper ``Status and
  Performance of the First VERITAS Telescope'' from these proceedings
} \presenter{Presenter: J.S. Perkins (jperkins@physics.wustl.edu), \
  usa-perkins-JS-abs-og23-oral}
\maketitle

\begin{abstract}
  We report on observations of the Perseus cluster of galaxies using
  the 10 m Whipple $\gamma$-ray telescope during the 2004-2005
  observing season.  We apply a two dimensional analysis technique to
  allow us to scrutinize the cluster for TeV emission.  In this
  contribution we will first calculate flux upper limits on TeV
  $\gamma$-ray sources within the cluster.  Second, we derive an upper
  limit on the extended cluster emission.  We then compare the flux
  upper limits with the EGRET upper limit at 100 MeV and theoretical
  models.
\end{abstract}

\section{Introduction}
We selected the Perseus cluster based on its closeness ($z\,=$ 0.0179,
75 Mpc) and high mass (4$\times10^{14}$~M$_\odot$~\cite{Giradi98}).
In galaxy clusters, $\gamma$-rays can originate from cosmic ray
electrons (CRe) as Inverse Compton and Bremsstrahlung emission and
from the decay of neutral pions produced when cosmic ray protons (CRp)
interact with thermal target material. Detection of $\gamma$-ray
emission from galaxy clusters would make it possible to measure the
energy density of these supra-thermal particles in the inter-cluster
medium. By combining MeV observations with TeV $\gamma$-ray
observations we may be able to disentangle the various components that
contribute to the emission \cite{Voelk99,Pfrommer04}. In addition to a
cosmic ray (CR) origin of $\gamma$-rays, annihilating dark matter may
also emit $\gamma$-rays. The intensity of the radiation depends on the
nature of dark matter, the annihilation cross sections, and the dark
matter density profile close to the core of the cluster
\cite{Horns04}.
\begin{figure}
\begin{minipage}{9cm}
\begin{center}
\includegraphics*[width=7cm,angle=0,clip]{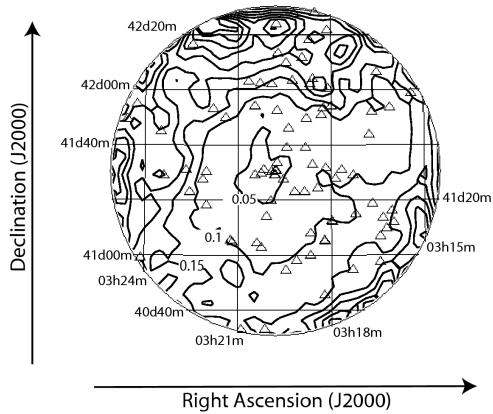}
\end{center}
\end{minipage}
\begin{minipage}{6cm}
\begin{center}
  \caption{\label {upper} 90\% upper limit map from point sources of
    the inner 1 degree of the Perseus cluster of galaxies.  The scale
    is in units of the flux from Crab Nebula with each contour step
    equal to 0.05 times the Crab flux.  Select contours are
    labeled. The triangles are radio sources, some of which can be
    found in Table~\ref{radio}.}
\end{center}
\end{minipage}
\end{figure}
\begin{table}   
\caption{\label{radio} $\gamma$-ray flux 90\% upper limits on the 5 most
  luminous radio galaxies in the Green Bank 6 cm Radio Source Catalog
  \cite{gregory96} within the Perseus cluster of galaxies.}
\begin{center}
\begin{tabular}{||c|c|c|c|c|c||} \hline \hline

~Source~  &   ~RA~  &  ~DEC~  & ~6 cm Flux~ & \multicolumn{2}{c||}{~TeV Flux Upper Limit~} \\ 
          & (J2000) & (J2000) &    (mJy)    & (Crab) & (ergs~cm$^{-2}$~s$^{-1}$) \\
                             \hline \hline

GB6 J0319+4130 & 03 19 47.1 & +41 30 42 & 46894& 0.047 & $0.37\times10^{-11}$\\
\hline
GB6 J0318+4153 & 03 18 16.0 & +41 53 14 & 15124& 0.10 & $0.78\times10^{-11}$\\
\hline
GB6 J0316+4118 & 03 16 40.9 & +41 18 49 & 264 & 0.14 & $1.1\times10^{-11}$\\
\hline
GB6 J0317+4054 & 03 17 19.4 & +40 54 40 & 53 & 0.19 & $1.5\times10^{-11}$\\
\hline
GB6 J0319+4223 & 03 19 44.4 & +42 23 24 & 49 & 0.34 & $2.6\times10^{-11}$\\
\hline
\end{tabular}
\end{center}
\end{table}

\section{Analysis and Results}
We observed the Perseus cluster of galaxies with the Whipple 10 m
telescope between August 16, 2004 and February 05, 2005 (UT). Details
about the Whipple telescope including the GRANITE-III camera have been
given in \cite{finley01}. Data from the Whipple 10 m are taken as
pairs of 28 minute runs. An ON run pointed at the source is followed
by an OFF run offset 7.50$^\circ$ in RA for background subtraction.
Removing runs with low rates and mismatched ON/OFF pairs results in a
usable data set of 29 ON/OFF pairs which corresponds to 810.4 minutes
of ON and OFF data. More detailed descriptions of Whipple observing
modes and analysis procedures can be found in
\cite{weekes96,punch91,reynolds93}.  The $\gamma$-ray selection
criteria used in this analysis (EZCuts2004,~see~\cite{kosack05}) were
designed to be independent of zenith angle and energy and are well
suited for two-dimensional source localization.  The 2D arrival
direction of each $\gamma$-ray event was calculated from the
orientation and elongation of the light distribution \cite{buckley98}.
We estimate a mean energy threshold for the Whipple 10 m data to be
approximately 400 GeV \cite{Rebillot05}. If not stated otherwise,
reported uncertainties are one standard deviation and upper limits are
given at the 90\% confidence level. The results given here are
preliminary.  Further analysis of systematic error is in progress.

In order to search for point sources within the field of view, the
camera resolution and efficiency need to be known to good accuracy at
all points.  We used an empirical method based upon data from the Crab
Nebula that was taken during the same months as Perseus. For the Crab
Nebula, we used 24 ON/OFF pairs with the camera centered on the Crab,
two pairs at an offset of 0.5 degrees and three pairs at an offset of
0.8 degrees resulting in a total on time of 670.7, 55.92 and 83.83
minutes respectively. The Crab data were binned by the square of the
distance of the reconstructed shower direction from the location of
the Crab Nebula (so as to eliminate any area scaling) and the excess
(ON minus OFF) was plotted and fitted with an exponential.  This fit
gives us a direct measurement of the resolution of the camera from a
point source at the three different offsets. From these same data we
determined an optimal angular cut based on the integral of the excess
as a function of the radius.  We also determined how the efficiency of
the camera falls off towards the edges by calculating the $\gamma$-ray
rate at the different offsets.

Using these results we then searched over the entire field of view of
the camera for point sources within the Perseus cluster.  At every
point on the camera, we used the optimal cut specified above and
calculated the $\gamma$-ray excess from the data.  From this excess,
we determined the flux in units of Crab.  We then used this flux and
its error to calculate a Bayes upper limit on the flux
\cite{Helene83}, taking into account the statistical error on the Crab
event rate.  See Figure \ref{upper} for the upper limit map.  Table
\ref{radio} shows the $\gamma$-ray flux upper limit for some of the
most luminous radio galaxies within the Perseus cluster from the Green
Bank 6 cm catalog \cite{gregory96}.

To search for extended emission from the Perseus cluster we assume
that the TeV $\gamma$-ray surface brightness mimics that of the
thermal X-ray emission (as seen from Chandra \cite{Sanders05} and
BeppoSAX \cite{Nevalainen04}) which arises from interactions of the
CRs with the thermal photons in the cluster.  The X-ray surface
brightness can be modeled \cite{Pfrommer04} as a double-$\beta$
profile given by
\begin{equation}
\label{king}
\Sigma(r) = (\sum_{i=1}^2 \Sigma_i^2(1+\frac{r^2}{r^2_i})^{-3\beta_i/2})^2
\end{equation}
where $\Sigma$ is the surface brightness, $\Sigma_i, r_i$ and
$\beta_i$ are model parameters found in \cite{Pfrommer04} and based on
data from \cite{Churazov03} and \cite{Struble99}.  

\begin{figure}
\begin{minipage}{9cm} 
\begin{center}
\includegraphics*[width=7cm,angle=0,clip]{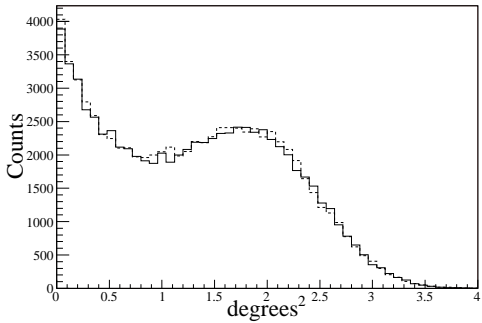}
\end{center}
\end{minipage}
\begin{minipage}{6.1cm}
\begin{center}
  \caption{\label {theta} Number of Whipple 10~m events versus the
    average distance of the estimated arrival direction from the
    center of the field of view squared.  The dashed data are the OFF
    counts and the solid are the ON counts.  There is good correlation
    between the ON and the OFF data out to the edge of the camera, and
    no excess from the cluster can be recognized.}
\end{center}
\end{minipage}
\end{figure}

The emission will continue out to the accretion shock which is
expected to occur at $\sim$2.2$^\circ$ from the cluster center.
Assuming the double-$\beta$ profile, we estimate that 84\% of the
total cluster emission comes from within 0.8$^\circ$ from the cluster
center.  Figure \ref{theta} shows the ON and OFF data after analysis
and cleaning plotted versus the distance from the center of the field
of view squared. There is no obvious excess in Figure \ref{theta} out
to the edge of the field of view.

We derive a quantitative upper limit by normalizing this profile,
smoothed with the Whipple angular resolution, to 1 over the field of
view of the camera.  We then smoothed the expected emission by the
resolution of the Whipple telescope and then multiplied the result by
the efficiency curve found from the Crab data to produce an expected
rate map. The rate map and actual excess were integrated over the
inner 0.8 degrees and these two values were used to determine the
upper limit on the TeV flux from the entire Perseus cluster.  The
upper limit on the emission is 0.23 of the Crab flux
($1.31\times10^{-11}$ ergs~cm$^{-2}$~s$^{-1}$).

\section{Interpretation and Discussion}
\begin{figure}
\begin{minipage}{9 cm}
\begin{center}
\includegraphics*[height=7cm,angle=270,clip]{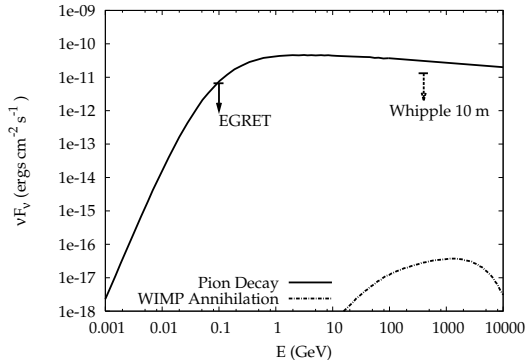}
\end{center}
\end{minipage}
\begin{minipage}{6cm}
\begin{center}
  \caption{\label {flux_predictions} The Whipple 90\% upper limit on
    the emission from the Perseus cluster is plotted as a dashed arrow
    at 400 GeV.  The EGRET upper limit from \cite{Reimer03} is shown
    at 100 MeV. The upper limits of the pion decay $\gamma$-ray flux
    from \cite{Pfrommer04} is plotted as a solid line. Also plotted is
    the dark matter emission derived under the assumption that the TeV
    $\gamma$-ray signal from the galactic center origionates from the
    annihilation of an 18 TeV neutralino \cite{Horns04}.}
\end{center}
\end{minipage}
\end{figure}
Figure \ref{flux_predictions} shows the upper limit on TeV emission
from the cluster and compares it to a previous upper limit from EGRET
\cite{Reimer03} and with the results of two model calculations.  The
solid line is a model of CRp induced $\gamma$-ray emission normalized
to the EGRET upper limit \cite{Pfrommer04}. The $\gamma$-ray emission
arises from the decay of neutral pions produced when CRp interact with
thermal target material. The model assumes that the CRp energy density
is 14\% of the thermal energy density.  The dotted line shows the
expected emission derived under the assumption that the TeV emission
from the galactic center
\cite{Aharonian04,Kosack04,Tsuchiya04,Horns04} originates from dark
matter annihilation and that the dark matter flux scales with total
mass and the inverse of the distance squared. The expected flux is
well below the sensitivity of present instruments. Comparison of the
upper limits with models of leptonic emission (Bremsstrahlung, Inverse
Compton) are in preparation.

Though we did not detect significant TeV $\gamma$-rays from the
Perseus cluster of galaxies in 841 minutes of observations, we are
able to determine two different types of upper limits on the emission:
we placed upper limits on the emission from point sources within the
cluster and we provide an overall upper limit assuming extended
emission. The $\gamma$-ray data constrain the ratio of energy of the
thermal and non-thermal plasmas.  

\section{Acknowledgments}
This research is supported by grants from the U.S. Department of
Energy, the National Science Foundation, the Smithsonian Institution,
by NSERC in Canada, by Science Foundation Ireland and by PPARC in the
UK.  J.S. Perkins would like to thank the AAS, the ICRC and the
Washington Univ. Physics Dept for travel support.

\end{document}